\documentclass[prl,aps,amssymb,twocolumn,showpacs]{revtex4} 

\usepackage{graphicx}
\usepackage{hypernat}
\usepackage{hyperref}
\usepackage{natbib}

\begin{document}
\pagebreak
\title{Microscopic Calculation of IBM Parameters
by Potential Energy Surface Mapping}
\author{I. Bentley$^1$}
\author{S. Frauendorf$^{1,2}$}
\affiliation{$^1$Dept. of Physics, University of Notre Dame, Notre Dame, IN 46556}
\affiliation{$^2$ISP, Forschungszentrum Dresden-Rossendorf, Dresden, Germany}

\date{\today}

\begin{abstract} 
A coherent state technique is used to generate an Interacting Boson Model (IBM) Hamiltonian energy surface which is adjusted to match a mean field energy surface. This technique allows for calculation of IBM Hamiltonian parameters,  prediction of properties of low lying collective states, as well as generation of probability distributions of various shapes in the ground state of transitional nuclei. The last two of which are of astrophysical interest. The results for krypton, molybdenum, palladium, cadmium, gadolinium, dysprosium and erbium nuclei are compared with experiment.
\end{abstract}


\maketitle
\section{Introduction}
The Interacting Boson Model (IBM) is a powerful tool for describing the low-lying collective quadrupole excitations \citep{JD71}, \citep{JJ74}. The IBM-1 formalism, in which protons and neutrons are indistinguishable, uses the approximation that pairs of nucleons behave like bosons with either angular momentum 0 or 2 \citep{AI75}. It leads to a class of particularly simple Hamiltonians \citep{SI78}, \citep{ZB02}, that include the consistent Q formalism \citep{WC82}, which will be the focus of the following discussion.

Conventionally, IBM Hamiltonians have been used to fit the experimental energy spectra and the electromagnetic transition probabilities. Within this framework, the Casten Triangle \citep{Ca06} can be used to classify the experimental spectra, which provides insight in terms of limiting symmetries as well as indicating phase transitions. 

Another application of the IBM is the Instantaneous Shape Sampling (ISS) technique, which has been used to calculate the $\gamma$-absorption strength functions of transitional nuclei \citep{BB11}, which are needed for studies of element synthesis in the cosmos. In the ISS, the IBM is used to generate the probability distribution of shapes in the ground state of transitional nuclei. These probability distributions can be used to calculate the deformation-dependent absorption probability of a photon. 

The parameters of the IBM-1 Hamiltonian have been obtained from fits to the experimental spectra. That is, the probability distribution, thus far, requires experimental input in order to determine the IBM parameters. However, in applying ISS to the r- and p- processes of element synthesis, one needs to predict the parameters of the IBM Hamiltonian for the nuclei involved. 

The calculation of the IBM Hamiltonian parameters from the underlying fermionic structure has remained a challenge, and predicting the parameters for nuclei far from stability is an even greater challenge. A new approach for calculating the IBM parameters has been suggested by Nomura et al. \citep{NS08}. The basic idea is to match a fermionic Potential Energy Surface (PES) $E_{MF}(\beta,\gamma)$ with the bosonic PES $E_{IBM}(\beta_B,\gamma_B)$. The authors demonstrated it is possible to use coherent states to match a PES generated with constraint Skyrme Hartree-Fock and BCS pairing with a bosonic PES generated from a IBM-2 Hamiltonian, in which protons and neutrons are treated separately. The  resulting levels adequately reproduced the development of the spectra from SU(5) to the SU(3) limit for $84\leq N\leq $100. 

The success of this endeavor comes as somewhat of a surprise, because the fermionic PES is commonly considered as a potential that must be complemented by a mass tensor in order to construct a collective Hamiltonian. Determining the parameters of the IBM Hamiltonian by PES matching fixes both the potential and the mass tensor, i.e. the IBM Hamiltonian implicitly correlates the mass tensor with the potential. It seems that the inherent symmetries of the IBM lead to a realistic relation between the potential and the mass tensor, which is not the case in the mean field approaches, such as the micro-macro model discussed below.
 
In this paper we modify the method of Nomura which uses the IBM-2 formalism. One parameter in the IBM-2 Hamiltonian $\chi$ for the protons was held constant for a given isotope chain based on experience from IBM phenomenology, leaving the value of $\chi$ for neutrons to be fit. Because the two $\chi$ values are added together in the Hamiltonian, this essentially reduces the IBM-2 to IBM-1. For this reason we have decided to adopt IBM-1. Further justification for use of the IBM-1 is the well established observation that the deformations of the protons and neutrons are approximately equal \citep{MZ04},\citep{MB99}. 

Conventionally, the energy scale remains a free parameter in the IBM. For this reason, we again take a somewhat different strategy than \citep{NS08}, whose energy scale resulting from matching absolute energies is too large to reproduce the experimental spectra in a systematic way. Following common IBM practice, we determine from the mean field PES only the parameters of the IBM-1 Hamiltonian that control the relative position of the levels, but not the overall energy scale. The energy scale is fixed by the energy of the first $2^+$ state, which is either taken from experiment or can be calculated by the cranking procedure described below. The same holds true for the scale of the E2 transition rates. 

As a further modification, we generate the fermionic PES by means of a micro-macro method. The relatively smooth change of the energy with the deformation parameters allows for setting up an automated fitting procedure. The Skyrme energy density functional used in \citep{NS08} generates potentials that are not smooth enough for automated fitting (see e.g. \citep{DG72}). 

\section{The IBM Hamiltonian}

The simple version of the IBM-1 Hamiltonian given in Refs. \citep{WC82}, \citep{MZ04} turned out to be well suited for our purposes. It contains two IBM parameters and the energy scale, such that:
\begin{equation}
\label{eqn:HIBM}
H_{IBM}(\zeta , \chi)= c_E\Big((1-\zeta) \hat{n}_d-\frac{\zeta}{4N_B} \hat{Q}^{\chi}\cdot\hat{Q}^{\chi}\Big),
\end{equation}
where $\hat{n}_d=d^{\dagger}\cdot\tilde{d}$, and $\hat{Q}^{\chi}=[s^{\dagger}\tilde{d}+d^{\dagger}s]^{(2)}+\chi[d^{\dagger}\tilde{d}]^{(2)}$. The creation operators for the two spins are denoted by $s^{\dagger}$ and $d^{\dagger}$ respectively \citep{IA87}. This provides a description of quadrupole states in even-even nuclei in terms of the SU(6) group. 
The Hamiltonian is diagonalized within the space of $N_B$ bosons. The number of bosons is taken to be half the number of valence nucleons in agreement with conventional IBM approaches. 

The ratios of the energy levels are determined by the IBM parameters $\chi$ and $\zeta$. The energy scale is determined by the parameter $c_E(\zeta ,\chi)$, which will be fixed by the energy of the first $2^+$ state. 

Table \ref{tbl:McCutchan} presents IBM parameters from previous work obtained by fitting the experimental energy ratios of gadolinium, dysprosium and erbium isotopes. 

The parameters $\zeta$ and $\chi$ define the Casten triangle within which most nuclei can be placed \citep{Ca06}. The U(5) vibrational limit corresponds to $\zeta=0$ and describes a spherical nucleus. The other limit, $\zeta=1$, corresponds to a well deformed nucleus. The O(6) limit corresponds to $\zeta=1$ and $\chi=0$, representing a nucleus that is instable with respect to the triaxiality parameter $\gamma$. The SU(3) oblate or prolate rotor limit is reached for $\zeta=1$ and $\chi=\frac{\sqrt{7}}{2}$ or $\chi=-\frac{\sqrt{7}}{2}$, respectively.

\begin{table}
\begin{center}
\caption{IBM $\chi$-$\zeta$ Parameters from Fitting $4^+_1/2^+_1$, $0_{\beta}/2^+_1$ and $2_{\gamma}/2^+_1$ Ratios by McCutchan et al. \citep{MZ04}. \label{tbl:McCutchan}}
\begin{tabular}{c|ccc|ccc|ccc}\toprule
$N$ & $^A$X & $\chi$ & $\zeta$ & $^A$X & $\chi$ & $\zeta$ & $^A$X & $\chi$ & $\zeta$ \\ 
86 & $^{150}$Gd & -1.32 & 0.30 & $^{152}$Dy & -1.10 & 0.35 & $^{154}$Er & -0.85 & 0.30 \\
88 & $^{152}$Gd & -1.32 & 0.41 & $^{154}$Dy & -1.09 & 0.49 & $^{156}$Er & -0.62 & 0.55 \\
90 & $^{154}$Gd & -1.10 & 0.59 & $^{156}$Dy & -0.85 & 0.62 & $^{158}$Er & -0.61 & 0.63 \\
92 & $^{156}$Gd & -0.86 & 0.72 & $^{158}$Dy & -0.67 & 0.71 & $^{160}$Er & -0.60 & 0.69 \\
94 & $^{158}$Gd & -0.80 & 0.75 & $^{160}$Dy & -0.49 & 0.81 & $^{162}$Er & -0.53 & 0.75 \\
96 & $^{160}$Gd & -0.53 & 0.84 & $^{162}$Dy & -0.31 & 0.92 & $^{164}$Er & -0.37 & 0.84 \\
98 & $^{162}$Gd & -0.30 & 0.98 & $^{164}$Dy & -0.26 & 0.98 & $^{166}$Er & -0.31 & 0.91 \\ \toprule
\end{tabular}
\end{center}
\end{table}

\section{The Fermionic Potential Energy Surface}

The mean field PES are generated by means of the Tilted-Axis Cranking (TAC) code described in \citep{Fr93}, with the rotational frequency set equal to zero. The TAC is a micro-macro mean field method, which allows for the calculation of the energy as a function of the deformation parameters, $\beta$ ($\varepsilon_2$) and $\gamma$. This combines a macroscopic deformed liquid drop with microscopic corrections for the pairing interaction and Strutinsky renormalization of levels based on a Nilsson potential. The pairing effects are calculated using standard Bardeen-Cooper-Schrieffer (BCS) model pairing based on the phenomenological fits by M\"{o}ller and Nix \citep{MN92}. The BCS pairing gaps used are a function of atomic mass number: 
\begin{equation}
\label{eqn:DBCS}
\Delta_{p}=\frac{13.4}{A^{1/2}} [MeV], \textrm{ and } \Delta_{n}=\frac{12.8}{A^{1/2}} [MeV].
\end{equation}

As discussed below, the TAC code also allows one to calculate the energy of the first $2^+$ state, which sets the energy scale as well as the scale of the $B(E2)$ values. 

The resulting deformation minima are generally consistent with experimentally determined deformations and the results of M\"{o}ller, Nix et al. \citep{MN95},\citep{MB08}. The PES generated for mapping include the quadrupole and triaxial degrees of freedom, with the hexadecapole optimized for each $\varepsilon_2$-$\gamma$ grid. The value used for the hexadecapole deformation parameter is determined using an automated minimization procedure, in which all three parameters are determined corresponding to the equilibrium deformation.

\section{The Mapping Procedure}

The resulting TAC PES will be compared with IBM energy surface until the closest match is found. The expectation value of the IBM-1 Hamiltonian with the coherent state, ($\big\vert N_B,\beta_B,\gamma_B \big>$) is used to create the IBM energy surface \citep{VC81}. The state is comprised as a product of boson creation operators ($\hat{B}^{+}$), with:
\begin{equation}
\label{Coherent}
\big\vert N_B,\beta_B,\gamma_B \big>=\frac{1}{\sqrt{N!}} (\hat{B}^+)^{N_B}\big\vert0\big>, 
\end{equation}
where
\begin{equation}
\hat{B}^{+}=s^++\beta_B\Big(cos(\gamma_B)d^+_0+\frac{sin(\gamma_B)}{\sqrt{2}}(d^+_2+d^+_{-2})\Big).
\end{equation}
The $d^+$ operators are coupled to angular momentum projections of 0 or $\pm$2.
This expectation value has been given by Ginoccio and Kirson \citep{GK80}:
\begin{displaymath}
 E_{IBM}(\beta_{B},\gamma_B)=\big< N_B,\beta_B,\gamma_B \big\vert H_{IBM} \big\vert N_B,\beta_B,\gamma_B \big> 
\end{displaymath}
\begin{displaymath}
=c_E \bigg(\frac{\frac{-5}{4}\zeta+\big((1-\zeta)N_B-\frac{1}{4}\zeta(1+\chi^2)\big)\big(\beta_{B}\big)^2}{1+(\beta_{B})^2}
\end{displaymath}
\begin{displaymath}
-\Big(\frac{\zeta(N_B-1)(\beta_{B})^2}{(1+(\beta_{B})^2)^2}\Big) 
\end{displaymath}
\begin{equation}
\times \Big(1-\sqrt{\frac{2}{7}}\chi \beta_{B} cos(3\gamma_B)+\frac{\chi^2}{14}(\beta_{B})^2\Big)\bigg).
\end{equation}

The bosonic deformation parameters are assumed to be related to the fermionic mean field deformation parameters by:
\begin{equation}
\label{eqn:cbeta}
\gamma_B=\gamma, \textrm{ and } \beta_{B}=c_{\beta} \varepsilon_2.
\end{equation}
Hence, there are four unknowns; two IBM parameters $\zeta$, $\chi$ and two scaling coefficients $c_E$, $c_{\beta}$. The $c_E$ is the total energy scale, and $c_{\beta}$ is the deformation scale. 

The parameters of the IBM-1 Hamiltonian are determined by means of the following procedure. First, the energy of the first $2^+$ state is fixed. In Table \ref{tbl:TabIBM} the $E(2_1^+)_{EXP}$ is used, but this can also be calculated by the TAC. For each combination of $\zeta,\chi$ parameters appearing in the process of fitting, the IBM-1 Hamiltonian is diagonalized, and the scale $c_E$ of the IBM energy surface is determined by the ratio:
\begin{equation}
\label{eqn:cE}
c_E(\zeta ,\chi)= \frac{E(2_1^+)_{EXP,TAC}}{E(2_1^+)_{IBM}(\zeta ,\chi)}.
\end{equation}

Simultaneously, the scale parameter $c_{\beta}$ is fixed by the requirement that the IBM energy surface has a minimum at $\beta_B(min)=c_{\beta} \varepsilon_2(min)$, where the TAC energy surface has a minimum at $\varepsilon_2(min)$. There is no such requirement in the $\gamma$ degree of freedom because (\ref{eqn:HIBM}) is not capable of creating a triaxial minimum. 

In addition, there is the effective boson charge $e_B$, which sets the scale for the reduced transition probabilities $B(E2)$. It is fixed by a ratio of $B(E2)$ values such that: 
\begin{equation}
\label{eqn:eB}
e_B^2(\zeta,\chi)= \frac{B(E2:2^+_1\rightarrow0^+_1)_{EXP,TAC}}{B(E2:2^+_1\rightarrow0^+_1)_{IBM}(\zeta ,\chi)}.
\end{equation}
We will focus primarily on using the experimental value $B(E2:2^+_1\rightarrow0^+_1)_{EXP}$. In regions where this state hasn't been measured experimentally, one should use the $B(E2:2^+_1\rightarrow0^+_1)_{TAC}$ calculated by means of the tidal wave method explained below. 

The energy surface generated from the IBM-1 Hamiltonian do not well reproduce those generated by the TAC over a wide range of energies. The wave functions of the low-lying states explore only the low-energy portion of the PES. Therefore, one should only map the low-energy part of the TAC PES. In our evaluation, it has been determined that best agreement with experimental spectra is obtained when regions of the TAC energy surface below 1 MeV are mapped to the IBM energy surface. 

The parameters $\zeta,\chi$ are found by minimizing the mean squared deviation $d^2$ between $E_{TAC}(\varepsilon_{2i},\gamma_i)$ and $E_{IBM}(\varepsilon_{2i},\gamma_i)$, 
\begin{equation}
d^2(\zeta,\chi, c_{\beta})=\sum_i \Big( E_{TAC}- E_{IBM}(\zeta, \chi, c_{\beta})\Big)^2,
\end{equation}
where the index $i$ is summed over all grid points for which the TAC PES is below 1 MeV.

Often the low-energy region of the TAC energy surface contain more structure than the corresponding IBM energy surface. Nevertheless, the IBM parameters obtained from the mapping procedure are shown in Table \ref{tbl:TabIBM}. 

\begin{table}
\begin{center}
\caption{Equilibrium Deformation Parameters Calculated by Means of a Micro-Macro Method and IBM Mapping Parameters based on $E(2_1^+)_{EXP}$ and $B(E2:2^+_1\rightarrow0^+_1)_{EXP}$. \label{tbl:TabIBM}}
\begin{tabular}{cccc|cccccc}\toprule
$^A$X & $\varepsilon_2$ & $\varepsilon_4$ & $\gamma$ & $N_B$ & $c_E$ & $c_{\beta}$ & $\chi$ & $\zeta$ & $e_B$ \\ 
$^{76}$Kr & -0.220 & 0.008 & 0 & 10 & 3.07 & 2.50 & 0.74 & 0.62 & 0.081 \\ 
$^{78}$Kr & -0.201 & 0.014 & 0 & 11 & 2.72 & 2.50 & 0.56 & 0.60 & 0.070 \\ 
$^{80}$Kr & 0.063 & 0.001 & 0 & 12 & 2.67 & 2.50 & 0.24 & 0.56 & 0.055 \\ 
$^{82}$Kr & 0.051 & 0.002 & 0 & 13 & 3.43 & 3.00 & -0.12 & 0.56 & 0.040 \\ 
$^{84}$Kr & 0.000 & 0.000 & 0 & 14 & 2.77 & 3.00 & -0.10 & 0.48 & 0.032 \\ 
& & & & & & & \\ 
$^{98}$Mo & 0.136 & -0.009 & 3 & 10 & 3.75 & 2.75 & -0.06 & 0.62 & 0.050 \\ 
$^{100}$Mo & 0.185 & -0.002 & 22 & 11 & 3.41 & 2.75 & -0.04 & 0.68 & 0.061 \\ 
$^{102}$Mo & 0.219 & 0.001 & 26 & 12 & 2.55 & 3.00 & -0.04 & 0.76 & 0.074 \\ 
$^{104}$Mo & 0.241 & 0.005 & 21 & 13 & 2.26 & 3.50 & -0.04 & 0.88 & 0.079 \\ 
$^{106}$Mo & 0.255 & 0.012 & 16 & 14 & 2.19 & 3.50 & -0.02 & 0.90 & 0.073 \\ 
$^{108}$Mo & -0.229 & 0.017 & 0 & 15 & 2.74 & 3.50 & 0.04 & 0.90 & 0.076 \\ 
& & & & & & & \\ 
$^{102}$Pd & 0.096 & -0.003 & 0 & 12 & 2.62 & 3.50 & -0.46 & 0.56 & 0.060 \\ 
$^{104}$Pd & 0.127 & 0.001 & 0 & 13 & 3.28 & 3.25 & -0.34 & 0.60 & 0.058 \\ 
$^{106}$Pd & 0.143 & 0.006 & 0 & 14 & 3.36 & 3.00 & -0.22 & 0.62 & 0.060 \\ 
$^{108}$Pd & 0.166 & 0.008 & 0 & 15 & 3.09 & 2.50 & -0.24 & 0.62 & 0.061 \\ 
$^{110}$Pd & 0.188 & 0.009 & 0 & 16 & 2.91 & 2.50 & -0.12 & 0.64 & 0.060 \\ 
$^{112}$Pd & 0.194 & 0.015 & 0 & 17 & 3.03 & 2.75 & -0.04 & 0.66 & 0.049 \\ 
$^{114}$Pd & -0.184 & 0.016 & 0 & 18 & 3.02 & 2.75 & 0.02 & 0.66 & 0.035 \\ 
$^{116}$Pd & 0.168 & 0.019 & 1 & 19 & 3.01 & 2.75 & 0.06 & 0.64 & 0.044 \\ 
& & & & & & & \\ 
$^{108}$Cd & 0.084 & 0.003 & 0 & 15 & 2.93 & 3.00 & -0.38 & 0.54 & 0.051 \\ 
$^{110}$Cd & 0.086 & 0.005 & 0 & 16 & 3.08 & 2.50 & -0.32 & 0.54 & 0.050 \\ 
$^{112}$Cd & 0.092 & 0.006 & 0 & 17 & 2.87 & 2.25 & -0.18 & 0.54 & 0.052 \\ 
$^{114}$Cd & -0.122 & -0.001 & 0 & 18 & 3.34 & 2.50 & -0.04 & 0.58 & 0.047 \\ 
$^{116}$Cd & -0.127 & 0.004 & 0 & 19 & 3.23 & 2.50 & 0.10 & 0.58 & 0.046 \\
& & & & & & & \\ 
$^{152}$Gd & 0.169 & -0.023 & 1 & 10 & 3.10 & 4.00 & -0.30 & 0.76 & 0.113 \\ 
$^{154}$Gd & 0.202 & -0.028 & 1 & 11 & 2.12 & 5.00 & -0.30 & 1.00 & 0.153 \\ 
$^{156}$Gd & 0.227 & -0.031 & 1 & 12 & 1.83 & 4.50 & -0.46 & 1.00 & 0.148 \\ 
$^{158}$Gd & 0.242 & -0.027 & 1 & 13 & 1.84 & 4.25 & -0.52 & 1.00 & 0.141 \\ 
$^{160}$Gd & 0.251 & -0.020 & 0 & 14 & 1.88 & 4.25 & -0.52 & 1.00 & 0.134 \\ 
& & & & & & & \\ 
$^{156}$Dy & 0.198 & -0.019 & 0 & 12 & 2.32 & 4.75 & -0.24 & 0.94 & 0.140 \\ 
$^{158}$Dy & 0.224 & -0.021 & 0 & 13 & 2.13 & 4.50 & -0.32 & 1.00 & 0.143 \\ 
$^{160}$Dy & 0.240 & -0.018 & 0 & 14 & 2.05 & 4.00 & -0.40 & 0.98 & 0.137 \\ 
$^{162}$Dy & 0.251 & -0.011 & 0 & 15 & 2.11 & 4.00 & -0.40 & 0.98 & 0.131 \\ 
$^{164}$Dy & 0.259 & -0.002 & 0 & 16 & 2.09 & 4.00 & -0.42 & 1.00 & 0.125 \\ 
& & & & & & & \\ 
$^{156}$Er & 0.155 & -0.014 & 0 & 12 & 2.97 & 3.75 & -0.26 & 0.70 & 0.100 \\ 
$^{158}$Er & 0.186 & -0.013 & 0 & 13 & 2.70 & 4.25 & -0.22 & 0.82 & 0.121 \\ 
$^{160}$Er & 0.215 & -0.013 & 0 & 14 & 2.55 & 4.50 & -0.20 & 0.96 & 0.133 \\ 
$^{162}$Er & 0.234 & -0.010 & 0 & 15 & 2.55 & 4.25 & -0.24 & 1.00 & 0.132 \\ 
$^{164}$Er & 0.248 & -0.003 & 0 & 16 & 2.46 & 4.25 & -0.28 & 1.00 & 0.128 \\ 
$^{166}$Er & 0.256 & 0.005 & 0 & 17 & 2.42 & 4.00 & -0.34 & 1.00 & 0.123 \\ 
$^{168}$Er & 0.261 & 0.014 & 0 & 18 & 2.55 & 4.00 & -0.36 & 1.00 & 0.116 \\ \toprule
 
\end{tabular}
\end{center}
\end{table}

Ref. \citep{FG10} describes the tidal wave approach that can be used to calculate the energy of the first $2^+$ state and the $B(E2)$ value, which fix $c_E$ and $e_B$ in regions where the experimental information is not available. To distinguish these results from the scales determined by experimental input, the TAC scales will be labeled as IBM Mapping with Calculated Scale (CS).

In the tidal wave approach, the yrast states $2^+$,$4^+$,... are viewed as a traveling wave that runs with a constant angular velocity over the surface of the nucleus. In the co-rotating frame, the time independent amplitude of this wave, i.e. deformation, can be calculated by means of the cranking model. The cranking model generates deformed states with $\big< J_x \big>=I(\omega)$. The energy $E_{TAC}(I(\omega),\varepsilon_2, \gamma)$ is minimized. The excitation energy between the states with $I=0$ and $2$ is
\begin{equation}
E(2_1^+)_{TAC}=E_{TAC}(I(\omega)=2)-E_{TAC}(I(\omega)=0).
\end{equation}

The TAC code has the advantage that it will generate the fermionic PES, $E(2_1^+)_{TAC}$ and $B(E2)_{TAC}$, all within the same frame work. The values obtained in this way for a number of the considered nuclides are given in \citep{FG10}. The resulting values typically agree with the experimental values within a range of 20\%. We have also carried out the mapping based on some calculated values. 

Table \ref{tbl:TACMAP} contains the resulting IBM parameters using the same micro-macro PES as before and but now taking the TAC generated $E(2_1^+)$ and $B(E2)$ to the set the scales in the mapping procedure. The procedure is now completely predictive. The IBM parameters can be compared to the results using the experimentally determined scales in Table \ref{tbl:TabIBM}. 

Eventhough the overall scales change, the $\chi$ and $\zeta$ parameters are only slightly modified, indicating that the initial fits to the potentials are somewhat robust. The resulting levels and transitions essentially contain a shifted scale depending on which approach is used.

\begin{table}
\begin{center}
\caption{IBM Mapping Parameters with CS based on $E(2_1^+)_{TAC}$ in [keV] and $B(E2:2^+_1\rightarrow 0^+_1)_{TAC}$ in [$e^2b^2$]. \label{tbl:TACMAP}}
\begin{tabular}{ccc|ccccc}\toprule
$^A$X & $E(2_1^+)_{TAC}$ & $B(E2)_{TAC}$  & $c_E$ & $c_{\beta}$ & $\chi$ &  $\zeta$ & $e_B$ \\
$^{98}$ Mo & 238 & 0.131 & 1.40 & 4.00 & -0.06 &  0.68 & 0.074 \\
$^{100}$ Mo & 178 & 0.202 & 1.45 & 3.75 & -0.08 &  0.76 & 0.082 \\
$^{102}$ Mo & 179 & 0.229 & 1.79 & 3.50 & -0.06 &  0.82 & 0.079 \\
$^{104}$ Mo & 155 & 0.278 & 1.96 & 4.00 & -0.02 &  0.96 & 0.080 \\
$^{106}$ Mo & 150 & 0.297 & 2.00 & 3.75 & -0.02 &  0.94 & 0.077 \\
$^{108}$ Mo & 149 & 0.199 & 2.40 & 4.00 & 0.04 &  1.00 & 0.060 \\
& & & & & & & \\ 
$^{102}$ Pd & 350 & 0.086 & 1.86 & 4.00 & -0.46 &  0.58 & 0.056 \\
$^{104}$ Pd & 311 & 0.113 & 2.12 & 3.75 & -0.36 &  0.62 & 0.058 \\
$^{106}$ Pd & 262 & 0.144 & 2.06 & 3.50 & -0.30 &  0.64 & 0.061 \\
$^{108}$ Pd & 247 & 0.154 & 2.02 & 3.00 & -0.26 &  0.64 & 0.060 \\
$^{110}$ Pd & 230 & 0.166 & 2.02 & 3.00 & -0.14 &  0.66 & 0.058 \\
$^{112}$ Pd & 224 & 0.175 & 2.13 & 3.00 & -0.06 &  0.68 & 0.055 \\
& & & & & & & \\ 
$^{108}$ Cd & 410 & 0.090 & 2.14 & 3.50 & -0.36 &  0.56 & 0.051 \\
$^{110}$ Cd & 397 & 0.080 & 2.11 & 3.00 & -0.30 &  0.56 & 0.046 \\
$^{112}$ Cd & 325 & 0.102 & 1.71 & 2.75 & -0.20 &  0.56 & 0.050 \\
$^{114}$ Cd & 270 & 0.117 & 1.81 & 3.25 & -0.02 &  0.60 & 0.048 \\ \toprule
\end{tabular}
\end{center}
\end{table}

\begin{figure}
\includegraphics[width=8.5cm]{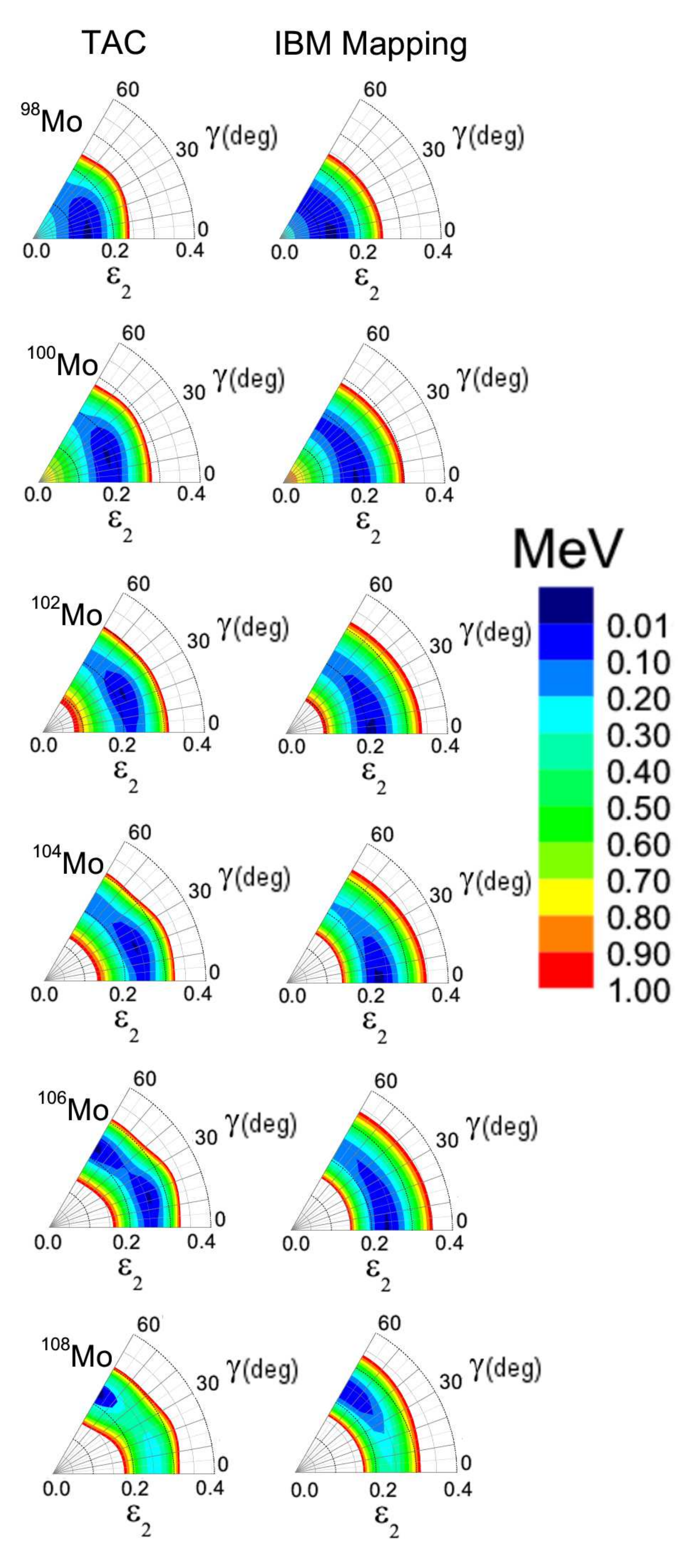}
\caption{(Color online) Energy surface comparisons for molybdenum nuclei, which are reproduced by the IBM. The color scale is in MeV, indicating the the region nearest the minima. The angle corresponds to the triaxial degree of freedom, with $\gamma=0$ corresponding to prolate shapes.} 
\label{fig:Fig.Mopots}
\end{figure}

\begin{figure}
\includegraphics[width=6.9cm]{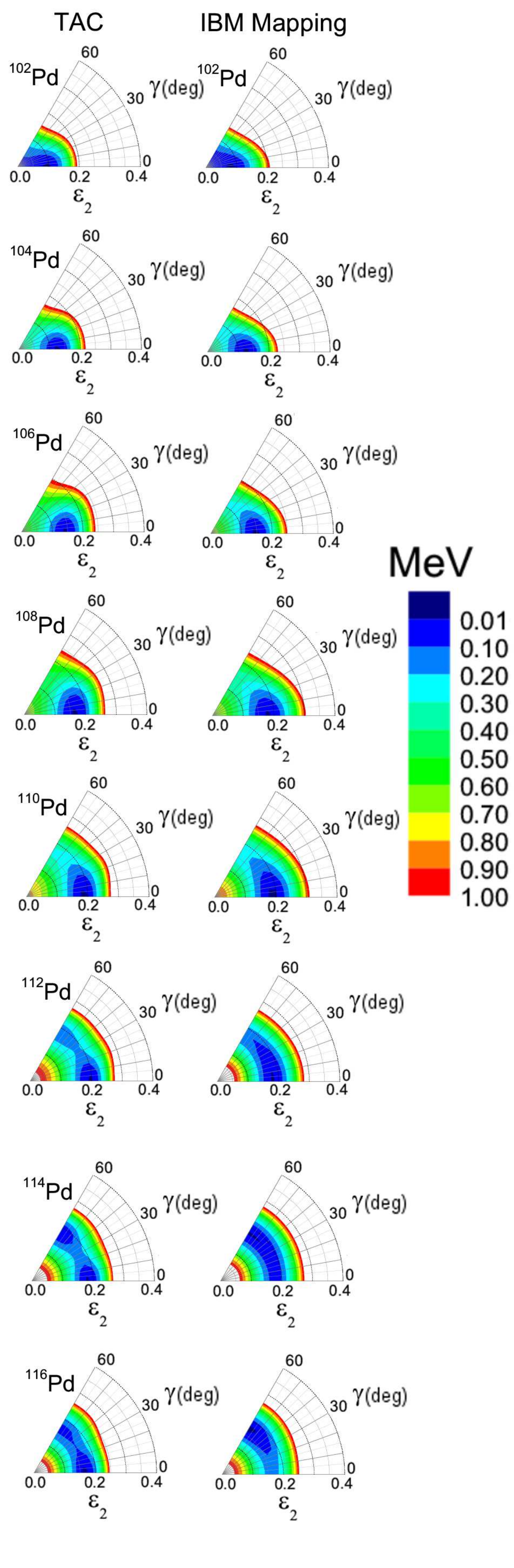}
\caption{(Color online) Mapping energy surface comparisons for palladium nuclei.} 
\label{fig:Fig.Pdpots}
\end{figure}

\begin{figure}
\includegraphics[width=8.5cm]{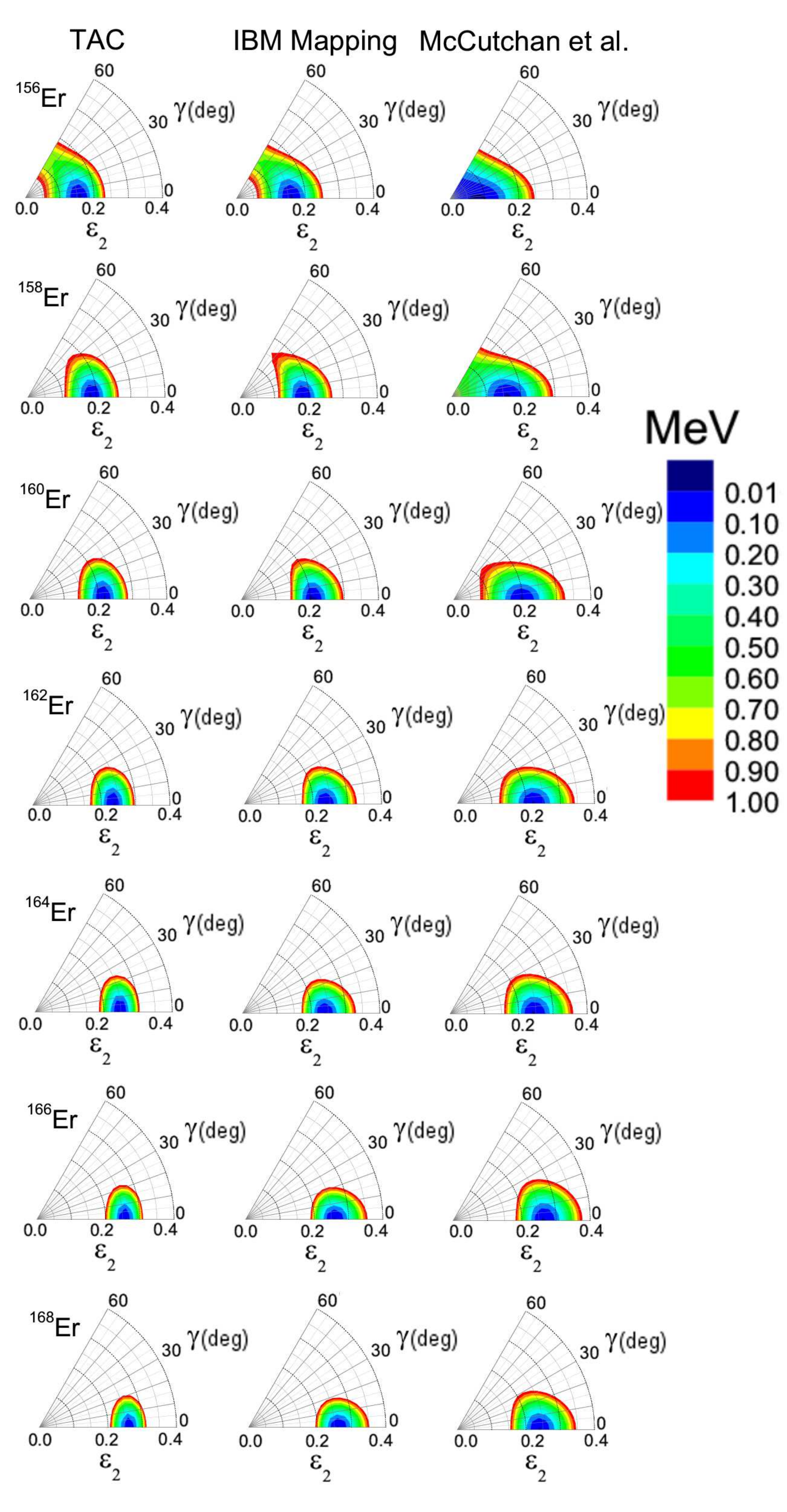}
\caption{(Color online) IBM energy surface comparisons for erbium nuclei. The third column contains the potentials created using McCutchan's $\chi$-$\zeta$ values, with $c_E$ determined by the experimental $2^+_1$ a constant scaling of $c_{\beta}=3.5$ } 
\label{fig:Fig.Erpots}
\end{figure}

Figures \ref{fig:Fig.Mopots}-\ref{fig:Fig.Erpots}, compare side by side the fermionic PES with the mapped bosonic energy surface. The color scale extends from the minimum up to 1 MeV above it, indicating the range used in the mapping procedure. The general features of the fermionic TAC energy surface can be reproduced by the mapped IBM energy surface.

The TAC energy surface of the molybdenum nuclei in Figure \ref{fig:Fig.Mopots} contain additional minima and triaxial minima which are not reproduced in the mapped IBM energy surface. This is a shortcoming of having used the simple Hamiltonian (\ref{eqn:HIBM}). However, these features are accounted for, to some extent, by a general $\gamma$-softening of the IBM energy surface. The $\gamma$-softness feature will be seen in the resulting levels as a lowering of the $\gamma$-band head. These results are similar to what was found for krypton.

Figure \ref{fig:Fig.Pdpots} contains the energy surface for palladium isotopes. For $^{112}$Pd-$^{116}$Pd a secondary prolate minima exists in the TAC PES, which eventually dominates. Similar results are found for cadmium nuclei. 

Figure \ref{fig:Fig.Erpots} compares the energy surface for erbium isotopes. The TAC energy surface with minima at large deformations ($\varepsilon_2 \geq$0.2) are not well reproduced by the IBM energy surface because it is not possible to create a large rigid deformation with the IBM-1 Hamiltonian that has been used. Similar results are found for the gadolinium and dysprosium nuclei. 

Previous IBM-1 fits to the experimental levels of the erbium isotopes by McCutchan et al. \citep{MZ04} indicate that the chain begins near the O(6) $\gamma$-soft limit and evolves toward the SU(3) rigid-rotor limit. The evolution of the parameters from mapping, in Table \ref{tbl:TabIBM}, begin in a different regime. Specifically, the values start close to the rotor limit, and slowly evolve toward the $\gamma$-soft limit.

Figure \ref{fig:Fig.Erpots} also contains the IBM energy surface generated from $\chi$-$\zeta$ values of McCutchan et al. where an average value was used for the scale $c_{\beta}$ of 3.5. The comparison of the parameters indicates an ambiguity in the mapping procedure. The ambiguity is that the low energy region of two energy surfaces with substantially different parameters can look nearly identical. Generally, reducing the value of $\zeta$ and increasing the value of $\chi$ leaves the PES nearly unchanged. This is the case when dealing with the erbium isotopes. 

The ambiguity doesn't persist far from the minima. For example, the potentials for $^{168}$Er at $\varepsilon_2=0$ are substantially different, $11.8 MeV$ and $4.6 MeV$, for the IBM Mapped parameters and McCutchan et al. fits.

\section{Energies and transition probabilities for selected nuclei}

\begin{figure*}
\includegraphics[width=13cm]{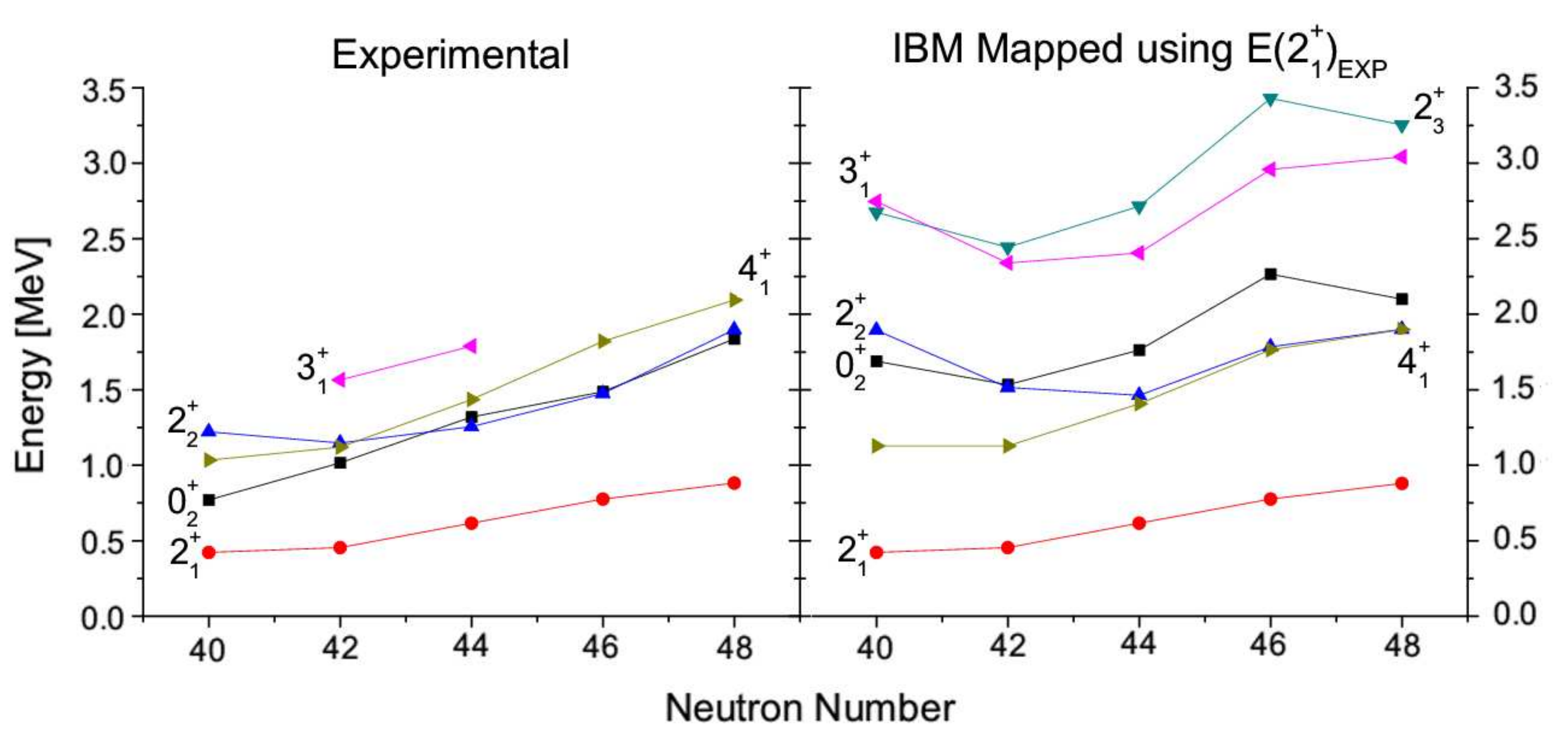}
\caption{(Color online) Low lying energy spectra for krypton nuclei with experimental data compared with results from IBM mapping techniques. Data retrieved from ENSDF \citep{Tu10}. Shape coexistence and the existence of a triaxial minima explain why the $\gamma$ and $\beta$-bands are lower than predicted. The $\gamma$-band refers to $E(2_3^+)$ and $E(3_1^+)$, while the $\beta$-band is $E(0_1^+)$ and $E(2_2^+)$.} 
\label{fig:Fig.Krlvls}
\end{figure*}

\begin{figure*}
\includegraphics[width=17.3cm]{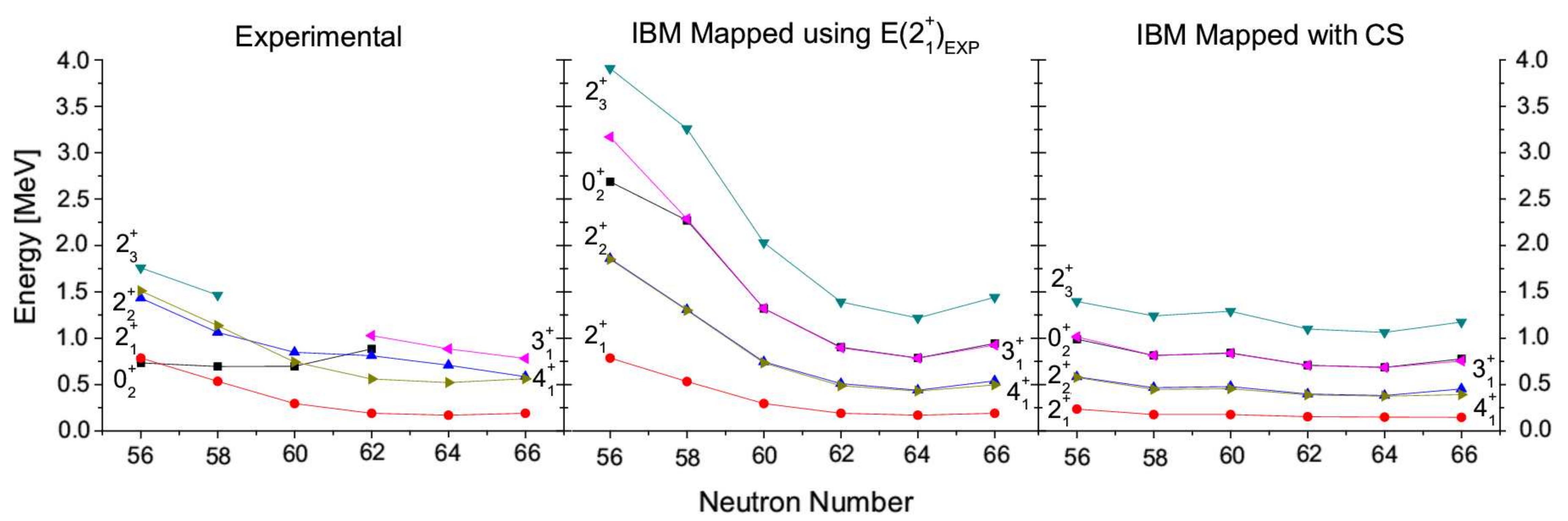}
\caption{(Color online) Low lying energy spectra for molybdenum nuclei experimental compared with IBM mapping results. Data retrieved from ENSDF \citep{Tu10}. The second and third columns consist of the mapping procedure using $E(2_1^+)_{EXP}$ and $E(2_1^+)_{TAC}$ to set $c_E$, respectively. }. 
\label{fig:Fig.Molvls}
\end{figure*}

\begin{figure*}
\includegraphics[width=17.3cm]{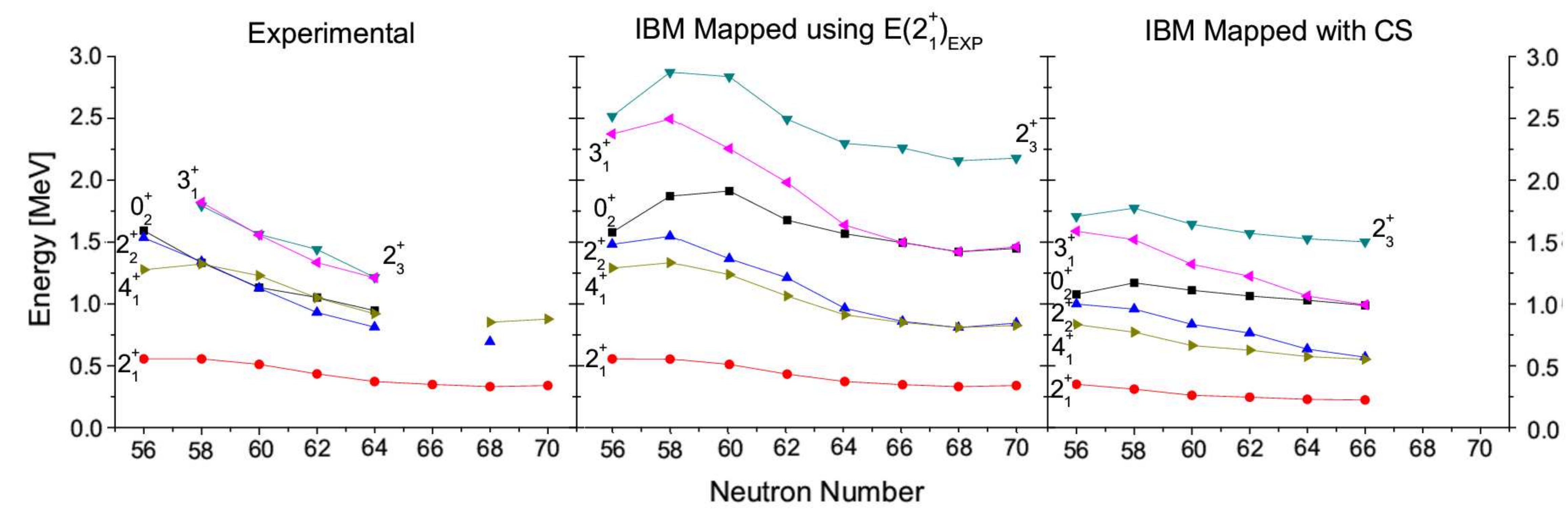}
\caption{(Color online) Low lying energy spectra for palladium nuclei experimental compared with IBM mapping results. Data retrieved from ENSDF \citep{Tu10}. The second and third columns are based on $E(2_1^+)_{EXP}$ and $E(2_1^+)_{TAC}$.} 
\label{fig:Fig.Pdlvls}
\end{figure*}

\begin{figure*}
\includegraphics[width=17.3cm]{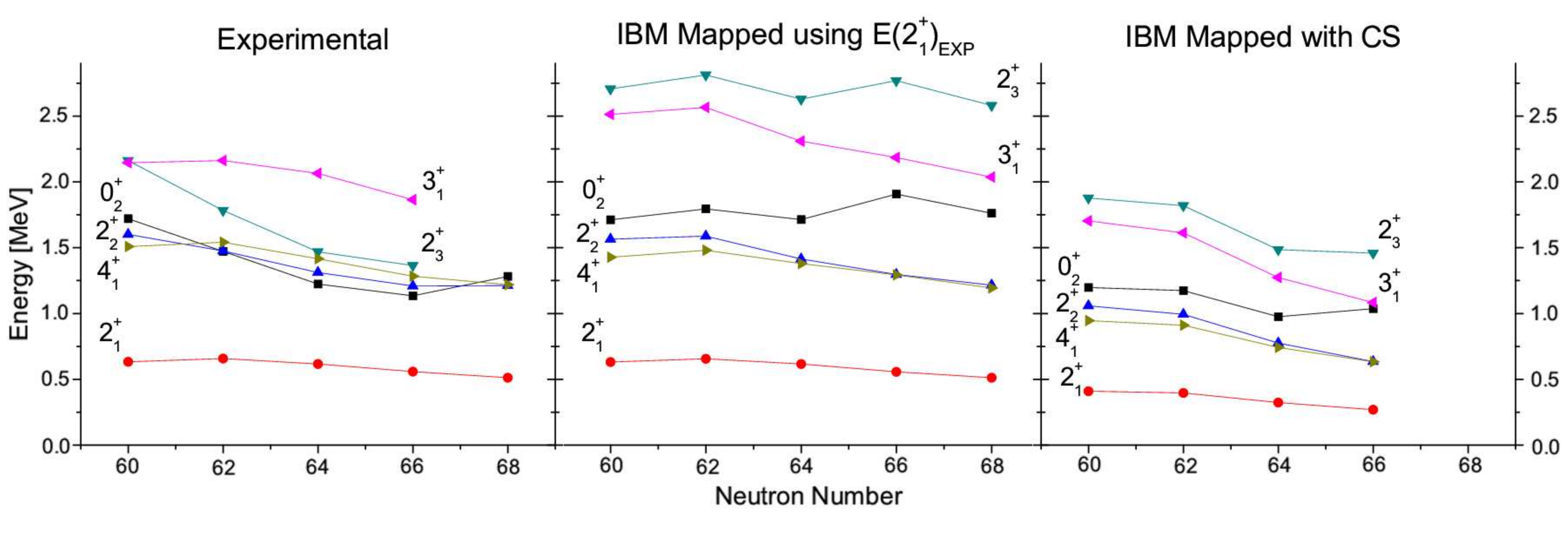}
\caption{(Color online) Low lying energy spectra for cadmium nuclei experimental compared with IBM mapping results. Data retrieved from ENSDF \citep{Tu10}. The second and third columns are based on $E(2_1^+)_{EXP}$ and $E(2_1^+)_{TAC}$.} 
\label{fig:Fig.Cdlvls}
\end{figure*}

\begin{figure*}
\includegraphics[width=17.3cm]{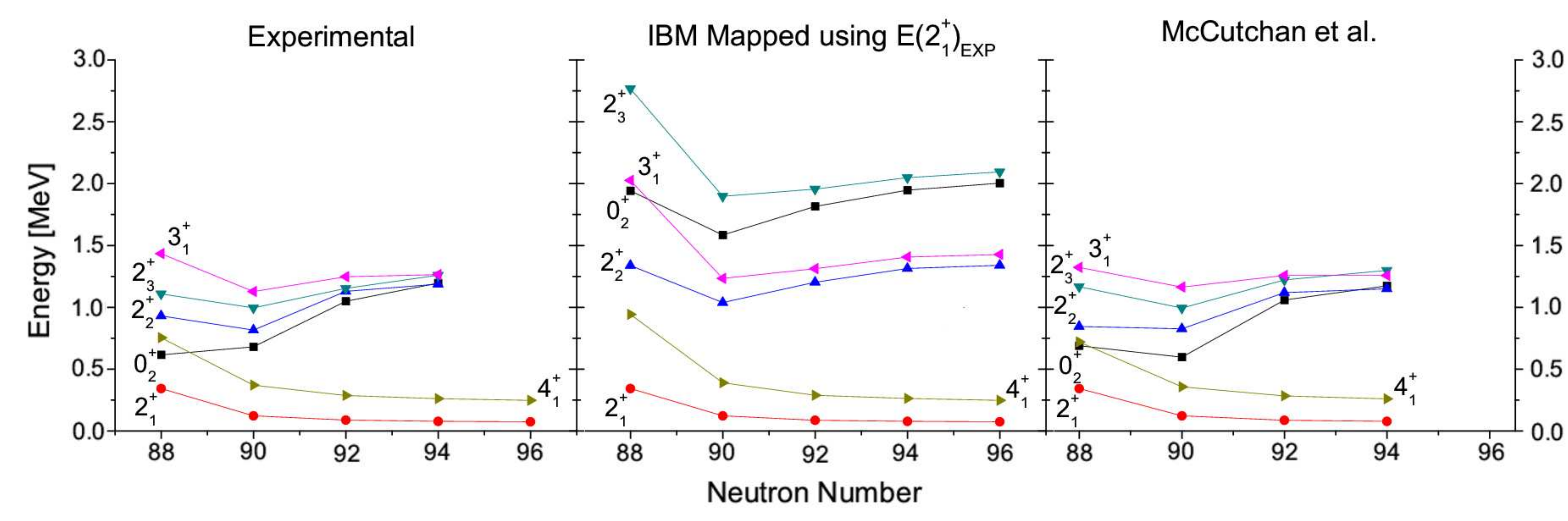}
\caption{(Color online) Low lying energy spectra for gadolinium nuclei experimental compared with IBM mapping results. Data retrieved from ENSDF \citep{Tu10}. Also included are the level spectra using the McCutchan et al. level fits \citep{MZ04}, which is scaled to the first experimental $2^+$.} 
\label{fig:Fig.Gdlvls}
\end{figure*}

\begin{figure*}
\includegraphics[width=17.3cm]{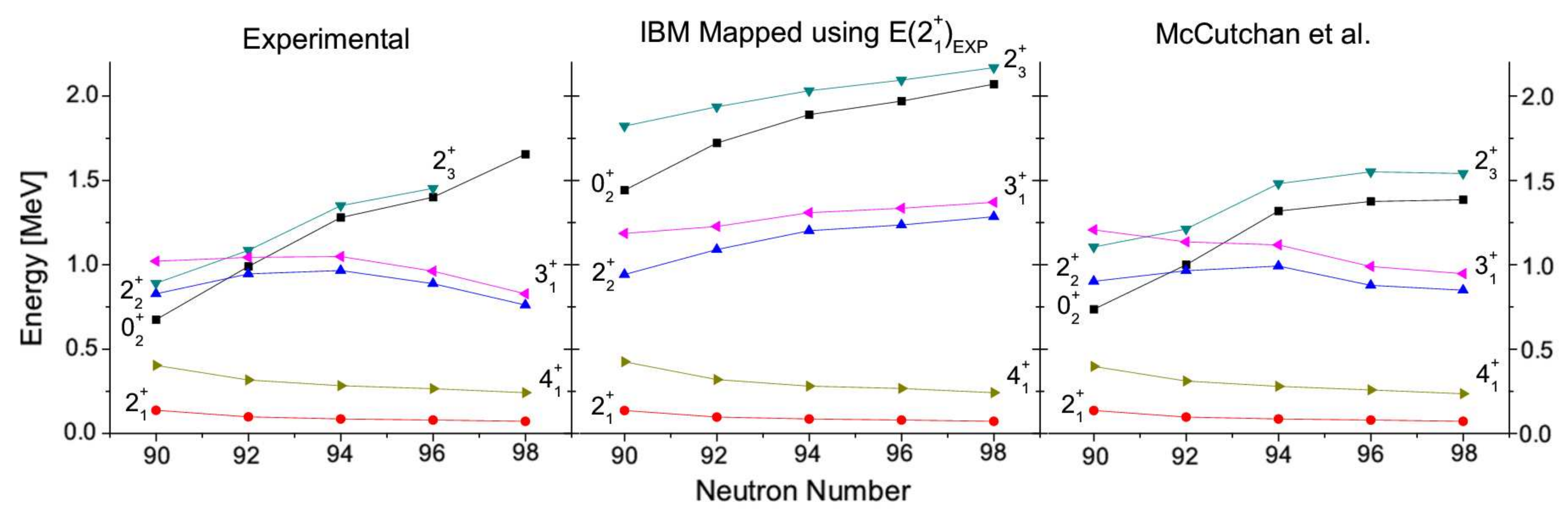}
\caption{(Color online) Low lying energy spectra for dysprosium nuclei experimental compared with IBM mapping results and the scaled level spectra using the level fits \citep{MZ04}. Experimental data retrieved from ENSDF \citep{Tu10}.} 
\label{fig:Fig.Dylvls}
\end{figure*}

\begin{figure*}
\includegraphics[width=17.3cm]{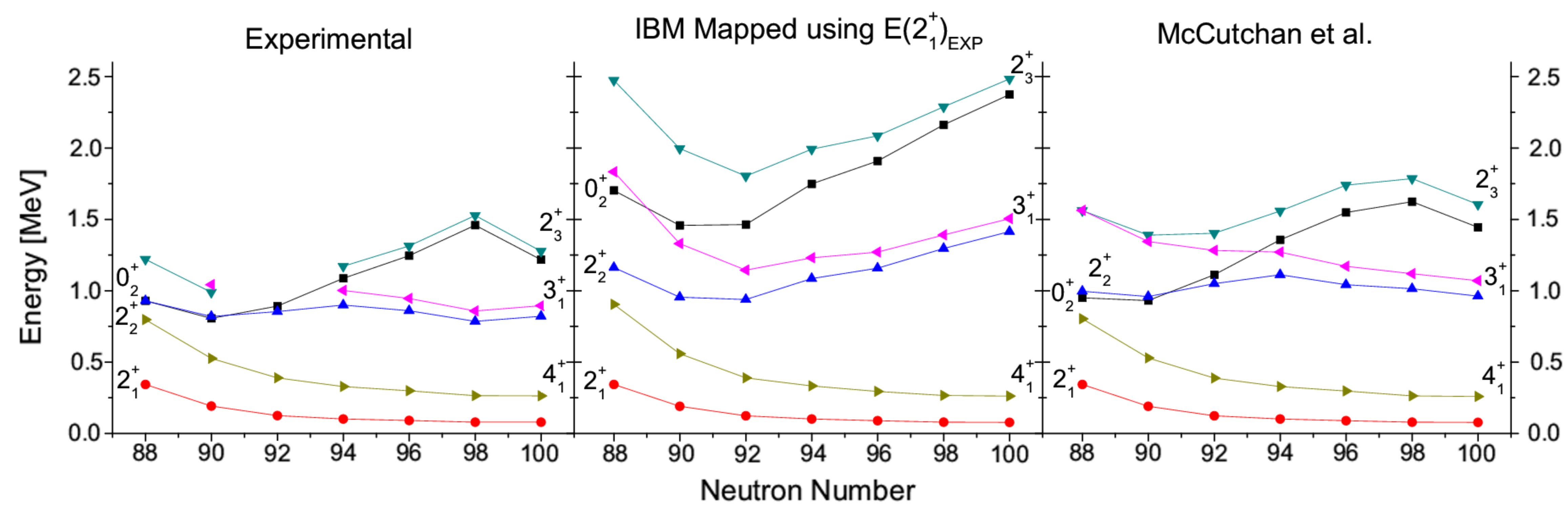}
\caption{(Color online) Low lying energy spectra for erbium nuclei experimental compared with IBM mapping results and the scaled level spectra using the level fits \citep{MZ04}. Experimental data retrieved from ENSDF \citep{Tu10}.} 
\label{fig:Fig.Erlvls}
\end{figure*}

The energies and transition probabilities that have been calculated by means of the IBM using parameters determined by the mapping procedure are shown in Figures \ref{fig:Fig.Krlvls}-\ref{fig:Fig.Erlvls} and Table \ref{tbl:TransitionTable} at the end of the text. The calculated values are compared with available experimental data. 

Figures \ref{fig:Fig.Krlvls} and \ref{fig:Fig.Molvls} show the levels of transitional and triaxial krypton and molybdenum nuclei. Although the energy surface are not very well reproduced, the resulting levels for the axial nuclei are in fair agreement with experiment. The exception is the second $0^+$, which lies too high in the calculation. Most likely this is the result of shape coexistence, which is indicated by the TAC energy surface but which the IBM Hamiltonian is unable to describe. As the neutron number increases, all of the results significantly improve.

Figures \ref{fig:Fig.Molvls}-\ref{fig:Fig.Cdlvls} contain a third column of levels calculated by IBM Mapping with CS that are completely predicted. These result from using the TAC to generate the transition and total energy scales. The corresponding spectra are basically very similar to those in the second column but are compressed based on the relative energies of the $2_1^+$ state.

Figures \ref{fig:Fig.Pdlvls} and \ref{fig:Fig.Cdlvls} display the palladium and cadmium isotopes. Their energy TAC energy surface are well reproduced by the mapped IBM energy surface. The ground state and $\gamma$-bands of palladium nuclei are within a few hundred keV of the experiment. For cadmium, the $4_1^+$ and $2_{\gamma}^+$ are in good agreement with experiment. Again, the $0^+_2$ states and the $2^+_3$ states built thereupon are least well accounted for. 

This is presumably related to the appearance of a secondary minimum in the TAC PES, which corresponds to the "intruder state" in IBM terminology. The first potential along the palladium isotopes which appears to have a second competing minima is $^{112}$Pd. As the neutron number increases for these isotopes, the secondary minimia becomes more apparent and the discrepancy between the mapped and experimental $\beta$-bands increase.

Figures \ref{fig:Fig.Gdlvls}-\ref{fig:Fig.Erlvls} include the strongly deformed isotopes of gadolinium, dysprosium and erbium, which have sharply rising energy surfaces that can only generated with many bosons $N_B>12$. The $\beta$-band is about 700 keV above the experimental values for dysprosium nuclei and 600 keV for erbium nuclei. 

In the well deformed nuclei, the $0^+_2$ states may contain an appreciable admixture of a pairing vibration or may represent a fragment of the $\beta$ vibration with a dominant two-quasiparticle contribution.

The transition probabilities, shown in Table \ref{tbl:TransitionTable}, are in reasonable agreement with experiment, except for the $0^+_2$ and $2^+_2$ states, for which the discrepancies are noticeably larger. Most of the transition rates for $^{106,108}$Pd and $^{112}$Cd are well reproduced by this technique. For $^{156}$Gd, even the weak transitions rates appear to be in agreement with experiment.

Overall, the most severe discrepancies between the calculations and experiment appear for the $0^+_2$ state, which experimental energy is below what is predicted. This is likely because the IBM Hamiltonian generates a $0^+_2$ state that has the character of a shape vibration. However, the structure of the $0^+_2$ states is more complex. 

Clearly, the IBM Hamiltonian does not account for these complexities. Another weakness of the IBM-1 Hamiltonian used, is that it is not capable of producing a triaxial minimum. Experimental evidence indicates that some of the studied nuclei change from oblate to triaxial and eventually to prolate shape as the number of neutrons increases \citep{ZB09}. Triaxial shapes result in experimental $2^+_{\gamma}$ levels which are lower than those produced by the IBM mapping procedure.

In spite of the moderate success of this approach to predict the low lying levels, the resulting probability distributions may yield more promising results. The potentials generated using different parameters are often comparable, take for example the erbium nuclei past $^{160}$Er in Figure \ref{fig:Fig.Erpots}. The resulting probability distributions based on these potentials should likewise be comparable.

\begin{figure*}
\includegraphics[width=17.3cm]{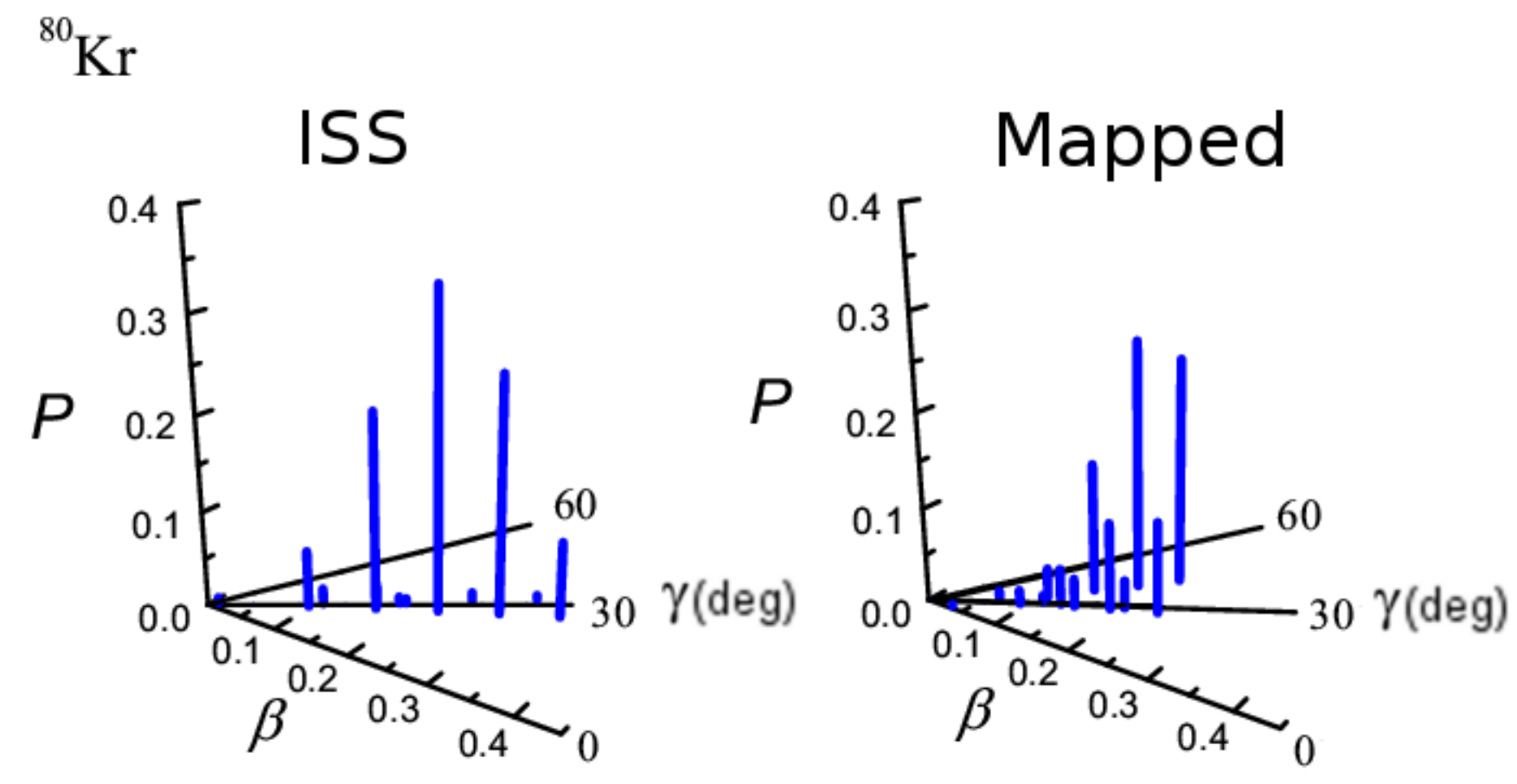}
\caption{(Color online) Probability distributions for the ground state of $^{80}$Kr based on the IBM mapping procedure and level fits from the ISS \citep{BB11}. For the ISS evaluation the IBM parameters are $\zeta=0.35$, $\chi=-0.06$, and $e_b=0.075$, while the mapped parameters are $\zeta=0.56$, $\chi=0.24$, and $e_b=0.055$.} 
\label{fig:Fig.80Kr}
\end{figure*}

The probability distributions for $^{80}$Kr determined by the experimental level fits in the ISS \citep{BB11} are compared with those resulting from the mapping procedure in Figure \ref{fig:Fig.80Kr}. In this case the value of $\zeta$ differed using the two approaches by $0.21$ and $\chi$ changed by $0.30$, however the probability distributions are similar. Furthermore, a photo-absorption cross-section calculation based on either of these probability distributions is expected to yield similar results.

\section{Conclusions}

We determined the parameters of the IBM-1 Hamiltonian by adjusting the bosonic PES, generated from the IBM Hamiltonian by calculating the expectation values of coherent states, to the fermionic PES, calculated by means of the micro-macro method. Matching the surfaces for energies less than 1 MeV above the minimum produced the best results. The overall energy scale was fixed by the energy of the $2^+_1$ state. 

The IBM-1 Hamiltonian, with parameters derived in this way reproduce the spectra of the even-even transitional and well deformed nuclei fairly well. The calculated energies of the ground state band and the quasi $\gamma$-band usually agree within a few hundred keV with experiment. The quasi $\beta$-band is not well reproduced, differing from experiment by up to 1 MeV. In almost all cases the calculated $0^+_2$ state lies higher than in experiment. We attribute this to the complex structure of this state. Shape coexistence, an admixture of pair vibrations, or a strong two-quasiparticle component, appear to modify its $\beta$ vibrational character. None of these effects are adequately accounted for by the IBM-1 Hamiltonian. 

The $B(E2)$ values for the ground band and the quasi $\gamma$-band are also reasonably well described. The calculated values for the quasi $\beta$-band may deviate substantially from experiment.

Some of the Mo isotopes have a triaxial minimum in their fermionic energy surface, which cannot be generated for bosonic energy surface derived from the IBM-1 Hamiltonian. In these cases the energy of the $2^+_2$ state is less well reproduced, as one might expect. 

It appears that the mapping technique can be used to predict the low-lying spectra of nuclei far from stability to the mentioned level of accuracy. In particular, the IBM parameters derived by mapping can be used for predicting the probability distributions of the ground state shapes of nuclei. This will allow for for the calculation of $\gamma$ absorption cross sections of nuclei that are unable to be measured experimentally, which is important for studying astrophysical processes. 

We would like to thank Piet Van Isacker for sharing the IBM-1 code that was used to calculate the nuclear levels and transition rates for a given set of IBM parameters. Additionally, we would like to thank Fritz D\"{o}nau for producing the probability distributions of $^{80}$Kr and Jie Sun for calculating the $E(2_1^+)_{TAC}$ and $B(E2:2^+_1\rightarrow 0^+_1)_{TAC}$ for molybdenum, palladium and cadmium. This work was supported by the DoE Grant DE-FG02-95ER4093.

\begin{table*}
\begin{center}
\caption{IBM Mapped Generated $B(E2)$ Transition Probabilities Compared to Experiment with Calculated Scale, Scaled to Match $B(E2:2^+_1\rightarrow 0^+_1)$ or $B(E2:2^+_2\rightarrow 0^+_1)$ from Experiment.\label{tbl:TransitionTable}}
\begin{tabular}{c|ccc|ccc|ccc|cc}\toprule
& \textbf{$^{106}$Pd} & & & \textbf{$^{108}$Pd} & & & \textbf{$^{112}$Cd} & & & \textbf{$^{156}$Gd}\\
$B(E2)$ & Exp. \citep{SF95} & Mapped & Mapped & Exp. \citep{SF95} & Mapped & Mapped & Exp. \citep{LG96} & Mapped & Mapped & Exp. \citep{Ap04} & Mapped\\
Transition & & $2^+_1\rightarrow 0^+_1$ & with CS & & $2^+_1\rightarrow 0^+_1$ & with CS & & $2^+_1\rightarrow 0^+_1$ & with CS & & $2^+_2\rightarrow 0^+_1$  \\
$0^+_2\rightarrow 2^+_1$ & 43$^{+6}_{-9}$ & 14.95 & 13.97 & 52$^{+6}_{-6}$ & 18.11 & 14.01 & 41$^{+9}_{-9}$ & 23.92 & 18.63 &  & 0.29 \\
$0^+_2\rightarrow 2^+_2$ & 19$^{+7}_{-3}$ & 31.62 & 39.1 & 47$^{+5}_{-11}$ & 40.74 & 46.9 &  & 6.71 & 9.96 & $>$5.4 & 29.89 \\
$0^+_3\rightarrow 2^+_1$ & 2.4$^{+0.4}_{-0.3}$ & 1.06 & 0.7 & $<$1 & 1.24 & 1.24 & 0.012$^{+0.001}_{-0.001}$ & 0.17 & 0.29 &  & 0.05 \\
$0^+_3\rightarrow 2^+_2$ & 13$^{+3}_{-2}$ & 39 & 28.63 & $<$19 & 42.58 & 29.25 &  & 54.74 & 44.38 &  & 5.05 \\
$2^+_1\rightarrow 0^+_1$ & 42$^{+4}_{-4}$ & 42 & 43.41 & 50$^{+7}_{-5}$ & 50 & 48.37 & 30.6$^{+0.6}_{-0.6}$ & 30.6 & 28.29 &  & 107.95 \\
$2^+_2\rightarrow 0^+_1$ & 0.87$^{+0.10}_{-0.09}$ & 0.82 & 1.51 & 0.63$^{+0.07}_{-0.07}$ & 1.22 & 1.63 &  & 0.15 & 0.25 & 2.7-4.2 & 3.45 \\
$2^+_2\rightarrow 2^+_1$ & 39$^{+4}_{-4}$ & 46.74 & 34.14 & 51$^{+5}_{-5}$ & 51.33 & 40.54 & 56$^{+25}_{-25}$ & 45.87 & 39.38 & 4.0-6.2 & 7.3 \\
$2^+_2\rightarrow 4^+_1$ &  & 0.01 & 0 &  & 0.01 & 0 &  & 0.14 & 0.11 & 0.6-1.0 & 0.6 \\
$2^+_3\rightarrow 0^+_1$ & 0.14$^{+0.02}_{-0.02}$ & 0.25 & 0.37 & 0.095$^{+0.010}_{-0.015}$ & 0.34 & 0.4 & 0.3$^{+0.1}_{-0.1}$ & 0.08 & 0.11 &  & 0.04 \\
$2^+_3\rightarrow 0^+_2$ & 39$^{+4}_{-4}$ & 28.2 & 28.2 & 59$^{+8}_{-6}$ & 33.84 & 31.68 & 59$^{+16}_{-16}$ & 25.11 & 22.12 &  & 76.06 \\
$2^+_3\rightarrow 0^+_3$ &  & 1.1 & 0.97 &  & 1.19 & 1.42 & 40$^{+20}_{-20}$ & 0.02 & 0.08 &  & 0.97 \\
$2^+_3\rightarrow 2^+_1$ & 0.52$^{+0.10}_{-0.07}$ & 0.03 & 0.06 & 1.7$^{+0.2}_{-0.7}$ & 0.05 & 0.07 & 0.3$^{+0.2}_{-0.2}$ & 0 & 0.01 &  & 0.03 \\
$2^+_3\rightarrow 2^+_2$ & 10.2$^{+2.2}_{-1.5}$ & 3.29 & 2.8 & 12$^{+6}_{-4}$ & 3.86 & 2.64 &  & 5.59 & 4.34 &  & 2.95 \\
$2^+_3\rightarrow 4^+_1$ & 5.3$^{+2.5}_{-1.4}$ & 6.74 & 6.12 & 45$^{+11}_{-7}$ & 8.11 & 6.38 &  & 10.01 & 7.88 &  & 0.15 \\
$2^+_4\rightarrow 0^+_2$ &  & 1.2 & 1.01 &  & 1.32 & 1.36 & 5.7$^{+1.5}_{-1.5}$ & 0.1 & 0.2 &  & 0.97 \\
$2^+_4\rightarrow 0^+_3$ &  & 26.64 & 24.22 &  & 31.45 & 28.76 & 26$^{+7}_{-7}$ & 24.79 & 21.49 &  & 73.96 \\
$2^+_4\rightarrow 2^+_2$ &  & 0.54 & 0.66 &  & 0.77 & 0.87 & $<$2.4 & 0.14 & 0.22 &  & 0.6 \\
$3^+_1\rightarrow 2^+_1$ &  & 1.46 & 2.71 &  & 2.18 & 2.91 & 1.7$^{+0.5}_{-0.5}$ & 0.29 & 0.47 & 6.1-9.1 & 6.04 \\
$3^+_1\rightarrow 2^+_2$ &  & 50.9 & 50.18 &  & 60.42 & 56.74 & 62$^{+17}_{-17}$ & 43.33 & 38.37 &  & 152.13 \\
$3^+_1\rightarrow 4^+_1$ &  & 17.16 & 13.86 &  & 19.38 & 16.38 & 24$^{+9}_{-9}$ & 16.46 & 14.11 & 4.1-6.1 & 5.03 \\
$4^+_1\rightarrow 2^+_1$ & 71$^{+7}_{-7}$ & 63.34 & 65.43 & 74$^{+5}_{-8}$ & 75.81 & 72.71 & 61$^{+8}_{-8}$ & 49.48 & 44.92 &  & 152.58 \\
$4^+_1\rightarrow 2^+_2$ &  & 0.01 & 0 & 1.2$^{+1.4}_{-1.1}$ & 0.01 & 0 &  & 0.08 & 0.06 &  & 0.33 \\
$4^+_2\rightarrow 2^+_1$ & 0.007$^{+0.006}_{-0.003}$ & 0.05 & 0.08 & $<$0.3 & 0.07 & 0.09 &  & 0.03 & 0.04 & 1.4-2.5 & 1.02 \\
$4^+_2\rightarrow 2^+_2$ & 35$^{+5}_{-4}$ & 37.77 & 37.05 & 55$^{+7}_{-5}$ & 44.84 & 41.84 &  & 31.99 & 28.4 &  & 54.7 \\
$4^+_2\rightarrow 2^+_3$ &  & 1.72 & 2.51 & 3.7$^{+2.7}_{-1.1}$ & 2.38 & 3.01 &  & 0.46 & 0.63 &  & 5.09 \\
$4^+_2\rightarrow 4^+_1$ & 23$^{+3}_{-2}$ & 27.73 & 21.87 & 30$^{+5}_{-5}$ & 31.12 & 25.78 &  & 27.39 & 23.35 & 8.2-14.6 & 8.4 \\
$4^+_3\rightarrow 2^+_1$ &  & 0 & 0.02 &  & 0.01 & 0.02 & 0.28$^{+0.08}_{-0.08}$ & 0 & 0 & 0 & 0.04 \\
$4^+_3\rightarrow 2^+_2$ &  & 0.39 & 0.86 & 2.9$^{+1.1}_{-0.8}$ & 0.62 & 0.88 & 69$^{+21}_{-21}$ & 0.11 & 0.16 & 1.7 & 0.47 \\
$4^+_3\rightarrow 2^+_3$ &  & 0.16 & 0.72 &  & 0.32 & 0.6 & 48$^{+14}_{-14}$ & 0.02 & 0.04 &  & 102.56 \\
$4^+_3\rightarrow 3^+_1$ &  & 39.68 & 36.19 &  & 46.62 & 41.58 &  & 36.5 & 31.83 & 3.7 & 0.1 \\
$4^+_3\rightarrow 4^+_1$ &  & 0.27 & 0.38 & 0.2$^{+1.5}_{-0.2}$ & 0.38 & 0.43 & 28$^{+9}_{-9}$ & 0.07 & 0.1 & 0.01 & 0 \\
$4^+_3\rightarrow 4^+_2$ &  & 34 & 30.67 & 1.9$^{+4.8}_{-1.8}$ & 39.82 & 35.9 &  & 31.07 & 26.99 & 2 & 5.94 \\
$4^+_3\rightarrow 6^+_1$ &  & 0.84 & 0.85 &  & 1.01 & 0.9 &  & 0.79 & 0.69 & 0 & 0.14 \\
$4^+_4\rightarrow 4^+_3$ &  & 1.54 & 3.64 &  & 2.46 & 3.44 &  & 0.44 & 0.63 & $>$35 & 24.96 \\
$5^+_1\rightarrow 4^+_1$ &  & 0.64 & 1.18 &  & 0.97 & 1.25 &  & 0.17 & 0.25 & 2.4-17.8 & 3.78 \\
$5^+_1\rightarrow 6^+_1$ &  & 16.09 & 13.55 &  & 18.47 & 15.97 &  & 15.61 & 13.35 & 3.6-22.6 & 6.66 \\
$6^+_1\rightarrow 4^+_1$ & 89$^{+10}_{-13}$ & 74.88 & 76.32 & 107$^{+12}_{-11}$ & 89.81 & 85.06 &  & 62.17 & 55.63 &  & 164.57 \\
$6^+_2\rightarrow 4^+_2$ &  & 54.88 & 55.04 & 56$^{+8}_{-17}$ & 65.94 & 61.98 &  & 47.91 & 42.34 &  & 105.44 \\
$8^+_1\rightarrow 6^+_1$ & 107$^{+13}_{-26}$ & 81.48 & 82.3 & 149$^{+19}_{-15}$ & 98.13 & 92.18 &  & 70.94 & 62.88 &  & 166.55 \\

\toprule
\end{tabular}
\end{center}
\end{table*}

\bibliography{bibliography}

\end{document}